\documentclass[pre,twocolumn,showpacs,showkeys,superscriptaddress,preprintnumbers,floatfix]{revtex4-1}

\usepackage{etex}
\usepackage{ifpdf}
\usepackage{hyperref}
\usepackage{dcolumn}
\usepackage{url}
\usepackage{amsmath}
\usepackage{amscd}
\usepackage{amsfonts}
\usepackage{amssymb}
\usepackage{bm}   
\usepackage{bbm}
\usepackage{verbatim}
\usepackage{stmaryrd}
\usepackage{amsthm}
\usepackage{xcolor}
\usepackage{setspace}
\usepackage{braket}

\usepackage{tikz}
\usetikzlibrary{arrows}
\usetikzlibrary{automata}
\usetikzlibrary{decorations.pathreplacing}
\usetikzlibrary{positioning}
\usetikzlibrary{plotmarks}
\usetikzlibrary{calc}
\usetikzlibrary{patterns}
\usetikzlibrary{external}
\usepackage{pgfplots}
\pgfplotsset{compat=newest}

\usepackage{graphicx}

\theoremstyle{plain}    
\theoremstyle{plain}    
\theoremstyle{plain}    
\theoremstyle{plain}    
\theoremstyle{plain}    
\theoremstyle{plain}    
\theoremstyle{plain}    
\theoremstyle{plain}    
\theoremstyle{plain}    
\theoremstyle{plain}    
\theoremstyle{plain}    
\theoremstyle{plain}

\newcommand{\eMs}    {\mbox{$\epsilon$-machines}}

\newcommand{\eTs}    {\mbox{$\epsilon$-transducers}}

\newcommand{\CausalState}   { \mathcal{S} }

\newcommand{\Cmu}       {C_\mu}
\newcommand{\hmu}       {h_\mu}
\newcommand{\EE}        {{\bf E}}

\newcommand{\forward}{+}
\newcommand{\reverse}{-}
\newcommand{\forwardreverse}{\pm} 

\newcommand{\FutureCausalState} { {\CausalState}^{\forward} }

\newcommand{\PastCausalState}   { {\CausalState}^{\reverse} }

\newcommand{\lastindex}[2]{
  \edef\tempa{0}
  \edef\tempb{#2}
  \ifx\tempa\tempb
    \edef\tempc{#1}
  \else
    \edef\tempa{0}
    \edef\tempb{#1}
    \ifx\tempa\tempb
      \edef\tempc{#2}
    \else
      \edef\tempc{#1+#2}
    \fi
  \fi
  \tempc
}

\newcommand{\I}{\mathbf{I}}

\newcommand{\CSjoint}[1][,]{
   \edef\tempa{:}
   \edef\tempb{#1}
   \ifx\tempa\tempb
      \ensuremath{\FutureCausalState\!#1\PastCausalState}
   \else
      \ensuremath{\FutureCausalState#1\PastCausalState}
   \fi
}

\newif\ifpm
\edef\tempa{\forwardreverse}
\edef\tempb{\pm}
\ifx\tempa\tempb
   \pmfalse
\else
   \pmtrue
\fi

\renewcommand{\H}{\operatorname{H}}
\renewcommand{\I}{\operatorname{I}}

\newcommand{\K}{\operatorname{K}}  
\newcommand{\kB}{k_\text{B}}  

\begin{document}

\title{Identifying Functional Thermodynamics\\
in\\
Autonomous Maxwellian Ratchets}

\author{Alexander B. Boyd}
\email{abboyd@ucdavis.edu}
\affiliation{Complexity Sciences Center and Physics Department,
University of California at Davis, One Shields Avenue, Davis, CA 95616}

\author{Dibyendu Mandal}
\email{dibyendu.mandal@berkeley.edu}
\affiliation{Department of Physics, University of California, Berkeley, CA
94720, U.S.A.}

\author{James P. Crutchfield}
\email{chaos@ucdavis.edu}
\affiliation{Complexity Sciences Center and Physics Department,
University of California at Davis, One Shields Avenue, Davis, CA 95616}

\date{\today}
\bibliographystyle{unsrt}

\begin{abstract}
We introduce a family of Maxwellian Demons for which correlations among
information bearing degrees of freedom can be calculated exactly and in compact
analytical form. This allows one to precisely determine Demon functional
thermodynamic operating regimes, when previous methods either misclassify or
simply fail due to approximations they invoke. This reveals that these Demons
are more functional than previous candidates. They too behave either as
engines, lifting a mass against gravity by extracting energy from a single heat
reservoir, or as Landauer erasers, consuming external work to remove
information from a sequence of binary symbols by decreasing their individual
uncertainty.  Going beyond these, our Demon exhibits a new functionality that
erases bits not by simply decreasing individual-symbol uncertainty, but by
increasing inter-bit correlations (that is, by adding temporal order) while
increasing single-symbol uncertainty. In all cases, but especially in the new
erasure regime, exactly accounting for informational correlations leads to
tight bounds on Demon performance, expressed as a refined Second Law of
Thermodynamics that relies on the Kolmogorov-Sinai entropy for dynamical
processes and not on changes purely in system configurational entropy, as
previously employed. We rigorously derive the refined Second Law under minimal
assumptions and so it applies quite broadly---for Demons with and without
memory and input sequences that are correlated or not. We note that general
Maxwellian Demons readily violate previously proposed, alternative such bounds,
while the current bound still holds.
\end{abstract}

\keywords{Maxwell's Demon, Maxwell's refrigerator, detailed balance, entropy rate, Second Law of Thermodynamics}

\pacs{
05.70.Ln  
89.70.-a  
05.20.-y  
05.45.-a  
}
\preprint{Santa Fe Institute Working Paper 15-07-025}
\preprint{arxiv.org:1507.01537 [cond-mat.stat-mech]}

\maketitle


\setstretch{1.1}
\section{Introduction}

The Second Law of Thermodynamics is only statistically true: while the entropy
production in any process is nonnegative on the average, $\langle \Delta S
\rangle \geq 0$, if we wait long enough, we shall see individual events for
which the entropy production is negative. This is nicely summarized in the
recent fluctuation theorem for the probability of entropy production
$\Delta S$~\cite{Evan93a, Evan1994, Gall95a, Kurc1998, Croo98a, Lebo1999, Coll2005}:
\begin{align}
  \frac{\Pr(\Delta S)}{\Pr(- \Delta S)} = e^{\Delta S}
  ~,
\label{eq:FT}
\end{align}	
implying that negative entropy production events are exponentially rare but not
impossible. Negative entropy fluctuations were known much before this modern
formulation. In fact, in 1867 J. C. Maxwell used the negative entropy
fluctuations in a clever thought experiment, involving an imaginary intelligent
being---later called Maxwell's Demon---that exploits fluctuations to
violate the Second Law~\cite{Leff02a, Maru2009}. The Demon controls a
small frictionless trapdoor on a partition inside a box of gas molecules to
sort, without any expenditure of work, faster molecules to one side and slower
ones to the other. This gives rise to a temperature gradient from an initially
uniform system---a violation of the Second Law. Note that the ``very observant
and neat fingered'' Demon's ``intelligence'' is necessary; a frictionless
trapdoor connected to a spring acting as a valve, for example, cannot achieve
the same feat~\cite{Smol16a}.

Maxwell's Demon posed a fundamental challenge. Either such a Demon could not
exist, even in principle, or the Second Law itself needed modification. A
glimmer of a resolution came with L. Szilard's reformulation of Maxwell's Demon
in terms of measurement and feedback-control of a single-molecule engine.
Critically, Szilard emphasized hitherto-neglected information-theoretic aspects
of the Demon's operations~\cite{Szil29a}. Later, through the works of R.
Landauer, O. Penrose, and C. Bennett, it was recognized that the Demon's
operation necessarily accumulated information and, for a repeating
thermodynamic cycle, erasing this information has an entropic cost that
ultimately compensates for the total amount of negative entropy production
leveraged by the Demon to extract work~\cite{Land61a, Penr70a, Benn82}. In
other words, with intelligence and information-processing capabilities, the
Demon merely shifts the entropy burden temporarily to an information reservoir,
such as its memory.  The cost is repaid whenever the information reservoir
becomes full and needs to be reset. This resolution is concisely summarized in
Landauer's Principle~\cite{Beru2012}: the Demon's erasure of one bit of
information at temperature $T \K$ requires at least $\kB T \ln{2}$ amount of
heat dissipation, where $\kB$ is Boltzmann's constant. (While it does not
affect the following directly, it has been known for some time that this
principle is only a special case \cite{Boyd14b}.)

Building on this, a modified Second Law was recently proposed that explicitly
addresses information processing in a thermodynamic system~\cite{Deff2013,
Bara2014a}:
\begin{align}
  \langle \Delta S \rangle + \kB \ln 2\, \Delta\!\H \geq 0
  ~,
\label{eq:SecondLaw}
\end{align}
where $\Delta\!\H$ is the change in the information reservoir's
configurational entropy over a thermodynamic
cycle. This is the change in the reservoir's ``information-bearing degrees of
freedom'' as measured using Shannon information $\H$~\cite{Cove06a}. These
degrees of freedom are coarse-grained states of the reservoir's
microstates---the mesoscopic states that store information needed for the
Demon's thermodynamic control.  Importantly for the following, this Second Law
assumes \emph{explicitly observed} Markov system dynamics \cite{Deff2013} and
quantifies this relevant information only in terms of the distribution of
\emph{instantaneous} system microstates; not, to emphasize, microstate path
entropies. In short, while the system's instantaneous distributions relax and
change over time, the information reservoir itself is not allowed to build up
and store memory or correlations.

Note that this framework differs from alternative approaches to the
thermodynamics of information processing, including: (i) active feedback control
by external means, where the thermodynamic account of the Demon's activities
tracks the mutual information between measurement outcomes and system
state~\cite{Touc2000, Cao2004, Saga2010, Toya10a, Ponm2010, Horo2010, Horo2011,
Gran2011, Abre2011, Vaik2011, Abre2012, Kund2012, Saga2012b, Kish2012}; (ii)
the multipartite framework where, for a set of interacting, stochastic
subsystems, the Second Law is expressed via their intrinsic entropy production,
correlations among them, and transfer entropy~\cite{Ito2013, Hart2014,
Horo2014, Horo2015}; and (iii) steady-state models that invoke time-scale
separation to identify a portion of the overall entropy production as an
information current~\cite{Espo2012, Stra2013}. A unified approach to these
perspectives was attempted in Refs.~\cite{Horo2013, Bara2014b, Horo2014b}.

Recently, Maxwellian Demons have been proposed to explore plausible
automated mechanisms that appeal to Eq.~(\ref{eq:SecondLaw})'s modified Second
Law to do useful work, by deceasing the physical entropy, at the expense of
positive change in reservoir Shannon information~\cite{Mand012a, Mand2013,
Stra2013, Bara2013, Hopp2014, Lu14a, Um2015}. Paralleling the modified Second
Law's development and the analyses of the alternatives above, they too neglect
correlations in the information-bearing components and, in particular, the
mechanisms by which those correlations develop over time. In effect, they
account for Demon information-processing by replacing the Shannon information
of the components as a whole by the sum of the components' \emph{individual}
Shannon informations. Since the latter is larger than the former
\cite{Cove06a}, using it can lead to either stricter or looser bounds than the
true bound which is derived from differences in total configurational
entropies. More troubling, though, bounds that ignore correlations can simply
be violated. Finally, and just as critically, they refer to configurational
entropies, not the intrinsic dynamical entropy over system trajectories.

This Letter proposes a new Demon for which, for the first time, all
correlations among system components can be explicitly accounted. This gives an
exact, analytical treatment of the thermodynamically relevant Shannon
information change---one that, in addition, accounts for system trajectories
not just information in instantaneous state distributions. The result is that,
under minimal assumptions, we derive a Second Law that refines
Eq.~(\ref{eq:SecondLaw}) by properly accounting for intrinsic information
processing reflected in temporal correlations via the overall dynamic's
Kolmogorov-Sinai entropy \cite{Dorf99a}.

Notably, our Demon is highly functional: Depending on model parameters, it acts
both as an engine, by extracting energy from a single reservoir and converting
it into work, and as an information eraser, erasing Shannon information at the
cost of the external input of work. Moreover, it supports a new and
counterintuitive thermodynamic functionality. In contrast with previously
reported erasure operations that only decreased single-bit uncertainty, we find
a new kind of erasure functionality during which multiple-bit uncertainties are
removed by adding correlation (i.e., by adding temporal order), while
single-bit uncertainties are actually increased. This new thermodynamic
function provocatively suggests why real-world ratchets support memory: The
very functioning of memoryful Demons relies on leveraging temporally correlated
fluctuations in their environment.

\section{Information Ratchets}

\begin{figure}[!ht]
\centering
\includegraphics[width=\columnwidth]{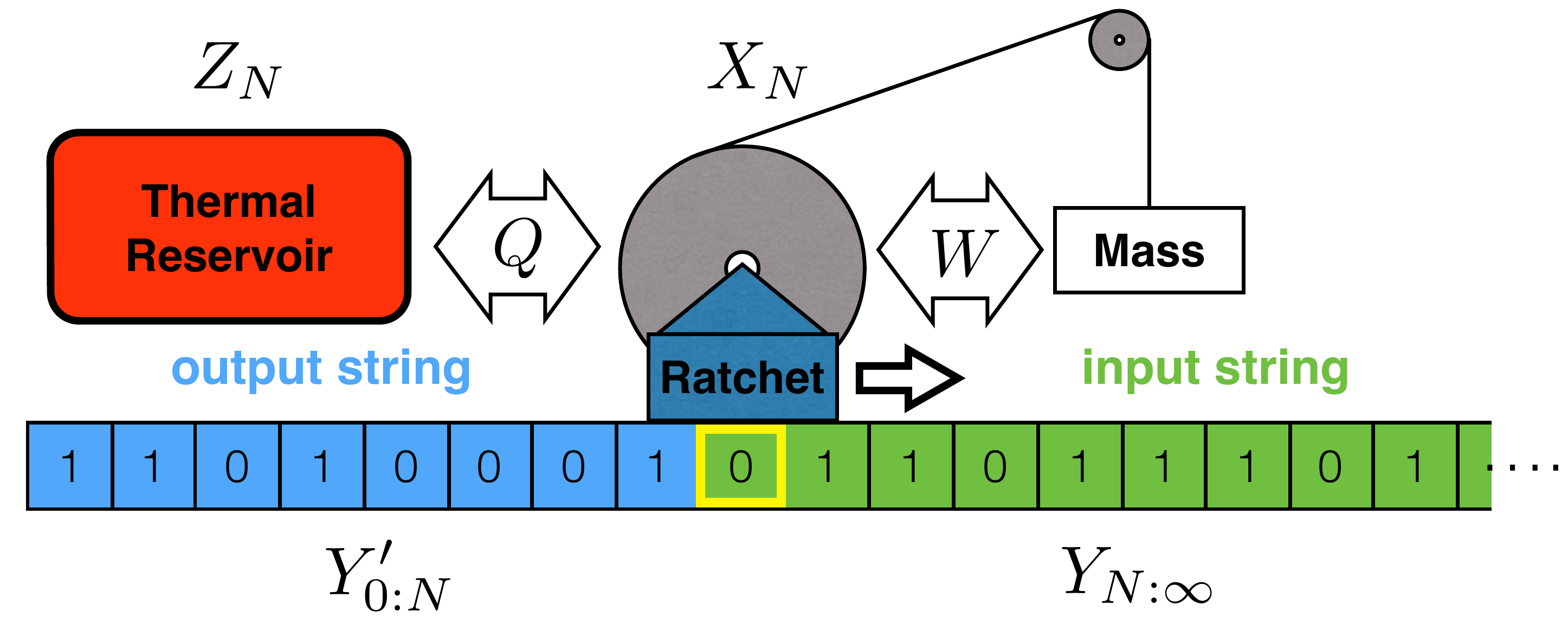}
\caption{Information ratchet sequentially processing a bit string: At time step
  $N$, $X_N$ is the random variable for the ratchet state and $Z_N$ that for
  the thermal reservoir. $Y_{N: \infty}$ is the block random variable for the
  input bit string and $Y'_{0:N}$ that for the output bit string. The last bit
  $Y_N$ of the input string, highlighted in yellow, interacts with the ratchet.
  The arrow on the right of the ratchet indicates the direction the ratchet moves along the tape as it
  sequentially interacts with each input bit in turn.
  }
\label{fig:FullRatchet}
\end{figure}

Our model consists of four components, see Fig. \ref{fig:FullRatchet}: (1) an
ensemble of bits that acts as an information reservoir; (2) a weight that acts
as a reservoir for storing work; (3) a thermal reservoir at temperature $T$;
and (4) a finite-state ratchet that mediates interactions between the three
reservoirs. The bits interact with the ratchet sequentially and, depending on
the incoming bit statistics and Demon parameters, the weight is either
raised or lowered against gravity.

As a device that reads and processes a tape of bits, this class of ratchet
model has a number of parallels that we mention now, partly to indicate
possible future applications. First, one imagines a sophisticated, stateful
biomolecule that scans a segment of DNA, say as a DNA polymerase does, leaving
behind a modified sequence of nucleotide base-pairs \cite{Albe14a}
or that acts as an enzyme sequentially catalyzing otherwise
unfavorable reactions \cite{Bril49a}. Second,
there is a rough similarity to a Turing machine sequentially recognizing tape
symbols, updating its internal state, and taking an action by modifying the
tape cell and moving its read-write head \cite{Lewi98a}. When the control logic
is stochastic, this sometimes is referred to as ``Brownian computing''
\cite[and references therein]{Stra15a}. Finally, we are reminded of the
deterministic finite-state tape processor of Ref. \cite{Moor90a} that, despite
its simplicity, indicates how undecidability can be imminent in dynamical
processes. Surely there are other intriguing parallels, but these give a sense
of a range of applications in which sequential information processing embedded
in a thermodynamic system has relevance.
	
The bit ensemble is a semi-infinite sequence, broken into incoming and outgoing
pieces. The ratchet runs along the sequence, interacting with each bit of the
input string step by step.  During each interaction at step $N$, the ratchet
state $X_N$ and interacting bit $Y_N$ fluctuate between different internal
joint states within $\mathcal{X}_N \otimes \mathcal{Y}_N$, exchanging energy with the thermal reservoir and work reservoir, and potentially changing $Y_N$'s state. At the end of step $N$, after input
bit $Y_N$ interacts with the ratchet, it becomes the last bit $Y'_{N}$ of the
output string. By interacting with the ensemble of bits, transducing the input
string into the output string, the ratchet can convert thermal energy from the
heat reservoir into work energy stored in the weight's height.
	
The ratchet interacts with each incoming bit for a time interval $\tau$,
starting at the $0$th bit $Y_0$ of the input string. After $N$ time intervals,
input bit $Y_{N-1}$ finishes interacting with the ratchet and, with the
coupling removed, it is effectively ``written'' to the output string, becoming
$Y'_{N-1}$. The ratchet then begins interacting with input bit $Y_N$. As Fig.
\ref{fig:FullRatchet} illustrates, the state of the overall system is described
by the realizations of four random variables: $X_N$ for the ratchet state,
$Y_{N:\infty}$ for the input string, $Y'_{0:N}$ for the output string, and
$Z_N$ for the thermal reservoir. A random variable like $X_N$ realizes elements
$x_N$ of its physical state space, denoted by alphabet $\mathcal{X}$, with
probability $\Pr(X_N=x_N)$. Random variable blocks are denoted $Y_{a:b} = Y_a
Y_{a+1} \ldots Y_{b-1}$, with the last index being exclusive. In the following,
we take binary alphabets for $\mathcal{Y}$ and $\mathcal{Y}'$: $y_N, y'_N \in
\{0,1\}$. The bit ensemble is considered two joint variables $Y'_{0:N} = Y'_0
Y'_1 \ldots Y'_{N-1}$ and $Y_{N: \infty} = Y_N Y_{N+1} \ldots$ rather than one
$Y_{0: \infty}$, so that the probability of realizing a \emph{word} $w\in
\{0,1\}^{b-a}$ in the output string is not the same as in the input string.
That is, during ratchet operation typically $\Pr(Y_{a:b}=w) \neq \Pr(Y'_{a:b}=w)$.

The ratchet steadily transduces the input bit sequence, described by the input
word distribution $\Pr(Y_{0:\infty})\equiv \{\Pr(Y_{0: \infty}=w)\}_{w \in
\{0,1\}^\infty}$---the probability for every
semi-infinite input word---into the output string, described by the word
distribution $\Pr(Y'_{0:\infty})\equiv\{\Pr(Y'_{0:\infty}=v)\}_{v \in
\{0,1\}^\ell}$. We assume that the word distributions we work with are
stationary, meaning that $\Pr(Y_{a:a+b}) = \Pr(Y_{0:b})$ for all nonnegative
integers $a$ and $b$.

A key question in working with a sequence such as $Y_{0:\infty}$ is how random it is. One
commonly turns to information theory to provide quantitative measures: the more
informative a sequence is, the more random it is. For words at a given length
$\ell$ the average amount of information in the $Y_{0:\infty}$ sequence is given by the
\emph{Shannon block entropy} \cite{Crut01a}:
\begin{align}
\H[Y_{0:\ell}] \equiv
	- \sum_{w \in \{0,1\}^\ell} \Pr(Y_{0: \ell}=w) \log_2 \Pr(Y_{0: \ell}=w)
  .
\end{align}
Due to correlations in typical process sequences, the irreducible randomness
per symbol is not the \emph{single-symbol entropy} $\H[Y_0]$. Rather,
it is given by the \emph{Shannon entropy rate} \cite{Crut01a}:
\begin{align}
\hmu \equiv \lim_{\ell \to \infty} \frac{\H[Y_{0:\ell}]}{\ell}
  ~.
\label{eq:EntropyRateViaBlocks}
\end{align}
When applied to a physical system described by a suitable symbolic dynamics, as done here, this quantity is the \emph{Kolmogorov-Sinai dynamical entropy} of the underlying physical behavior.

Note that these ways of monitoring information are quantitatively quite
different. For large $\ell$, $\hmu \ell \ll \H[Y_{0:\ell}]$ and, in particular,
anticipating later use, $\hmu \leq \H[Y_{0}]$, typically much less. Equality
between the single-symbol entropy and entropy rate is only achieved when the
generating process is memoryless.  Calculating the single-symbol entropy is
typically quite easy, while calculating $\hmu$ for general processes has been
known for quite some time to be difficult \cite{Blac57b} and it remains a
technical challenge \cite{Marc11a}. The entropy rates of the output sequence
and input sequence are $\hmu' = \lim_{\ell \to \infty}
\H[Y'_{0:\ell}]/\ell$ and $\hmu = \lim_{\ell \to \infty}
\H[Y_{0:\ell}]/\ell$, respectively.

The informational properties of the input and output word distributions set
bounds on energy flows in the system. Appendix \ref{app:NewSecondLaw}
establishes one of our main results: The average work done by the ratchet is
bounded above by the difference in Kolmogorov-Sinai entropy of the input
and output processes \footnote{Reference \cite{Mand012a}'s appendix suggests
Eq. (\ref{eq:NewSecondLaw}) without any detailed proof. An integrated version appeared also in Ref. \cite{Merh15a} for the special case of memoryless demons. 
Our App. \ref{app:NewSecondLaw} gives a more general proof of Eq.
(\ref{eq:NewSecondLaw}) that, in addition, accounts for memory.}:
\begin{align}
\langle W \rangle & \leq \kB T \ln 2 \, (\hmu' - \hmu) \nonumber \\
                  & = \kB T \ln 2 \, \Delta \hmu
   ~.
\label{eq:NewSecondLaw}
\end{align}
In light of the preceding remarks on the basic difference between $\H[Y_0]$ and
$\hmu$, we can now consider more directly the differences between Eqs.
(\ref{eq:SecondLaw}) and (\ref{eq:NewSecondLaw}). Most importantly, the $\Delta
\!\H$ in the former refers to the instantaneous configurational entropy $\H$
before and after a thermodynamic transformation. In the ratchet's steady state
operation, $\Delta \!\H$ vanishes since the configuration distribution is time
invariant, even when the overall system's information production is positive.
The entropies $\hmu^\prime$ and $\hmu$ in Eq. (\ref{eq:NewSecondLaw}), in
contrast, are dynamical: rates of active information generation in the input
and output giving, in addition, the correct minimum rates since they take all
temporal correlations into account. Together they bound the overall system's
information production in steady state away from zero. In short, though often
conflated, configurational entropy and dynamical entropy capture two very
different kinds of information and they, per force, are associated with
different physical properties supporting different kinds of information
processing. They are comparable only in special cases.

For example, if one puts aside this basic difference to facilitate comparison and considers the Shannon
entropy change $\Delta \!\H$ in the joint state space of all bits, the two
equations are analogous in the current setup. However, often enough, a weaker
version of Eq. (\ref{eq:SecondLaw}) is considered in the discussions on
Maxwell's Demon \cite{Mand012a,Mand2013,Bara2013,Bara2014b,Merh15a} and
information reservoirs \cite{Bara2014a}, wherein the statistical correlations
between the bits are neglected, and one simply interprets $\Delta \H$ to be the
change in the marginal Shannon entropies $\H[Y_{0}]$ of the individual bits. This implies
the following relation in the current context:
\begin{align}
\langle W \rangle \leq \kB \ln 2\, \Delta\!\H[Y_{0}]
  ~,
\label{eq:NotFundamental}
\end{align}
where $\Delta \H[Y_{0}] = \H[Y_0^\prime] - \H[Y_0]$. While Eq. (\ref{eq:NotFundamental}) is valid
for the studies in Refs.
\cite{Mand012a,Mand2013,Bara2013,Bara2014b,Bara2014a,Merh15a}, it cannot be
taken as a fundamental law, because it can be violated \cite{Chap15a}. In
comparison, Eq. (\ref{eq:NewSecondLaw}) is always valid and can even provide a stronger bound.

As an example, consider the case where the ratchet has memory and, for
simplicity of exposition, is driven by an uncorrelated input process, meaning
the input process entropy rate is the same as the single-symbol entropy: $\hmu
= \H[Y_0]$. However, the ratchet's memory can create correlations in the output
bit string, so:
\begin{align}
\Delta \hmu & = \hmu' - \H[Y_0] \nonumber \\
             & \leq \H[Y'_0] - \H[Y_0] \nonumber \\
			 & = \Delta\!\H[Y_{0}]
			 ~.
\end{align}
In this case, Eq. (\ref{eq:NewSecondLaw}) is a tighter bound on the work done
by the ratchet---a bound that explicitly accounts for correlations within the
output bit string the ratchet generates during its operation. For example, for
the combination $\{p = 0.5, q = 0.1, b = 0.9\}$, two bits in the outgoing
string are correlated even when they are separated by $13$ steps. Previously,
the effect of these correlations has not been calculated, but they have
important consequences. Due to correlations, it is possible to have an increase
in the single-symbol entropy difference $\Delta H[Y_0]$ but a decrease in the
Kolmogorov-Sinai entropy rate $\Delta h_\mu$. In this situation, it is
erroneous to assume that there is an increase in the information content in the
bits. There is, in fact, a decrease in information because of the correlations;
cf. Sec. \ref{sec:TF}.

Note that a somewhat different situation was considered in Ref.
\cite{Merh15a}, a memoryless channel (ratchet) driven by a correlated process.
In this special case---ratchets unable to leverage or create temporal
correlations---either Eq. (\ref{eq:NotFundamental}) or Eq.
(\ref{eq:NewSecondLaw}) can be a tighter quantitative bound on work. When a
memoryless ratchet is driven by uncorrelated input, though, the bounds are
equivalent. Critically, for memoryful ratchets driven by correlated input Eq.
(\ref{eq:NotFundamental}) can be violated. In all settings, Eq.
(\ref{eq:NewSecondLaw}) holds.

While we defer it's development to a sequel, Eq. (\ref{eq:NewSecondLaw}) also
has implications for ratchet functioning when the input bits are correlated as
well. Specifically, correlations in the input bits can be leveraged by the
ratchet to do additional work---work that cannot be accounted for if one only
considers single-symbol configurational entropy of the input bits
\cite{Riec15a}.

\section{Energetics and Dynamics}

\begin{figure}[tbp]
\includegraphics[width=\columnwidth]{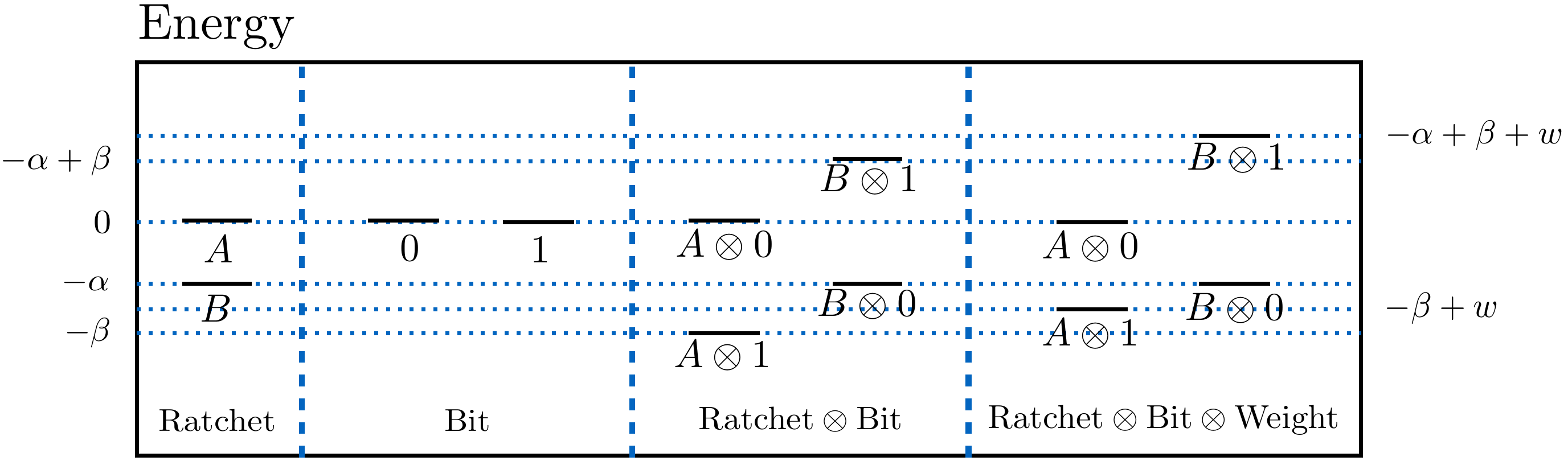}
\caption{Energy levels of the Demon states, interacting bits, their joint
  system, and their joint system with a weight in units of $[\kB T]$.
  }
\label{fig:Energies}
\end{figure}

To predict how the ratchet interacts with the bit string and weight, we need to
specify the string and ratchet energies. When not interacting with the ratchet
the energies, $E_0$ and $E_1$, of both bit states, $Y = 0$ and $Y = 1$, are
taken to be zero for symmetry and simplicity: $E_0 = E_1 = 0$. For simplicity,
too, we say the ratchet mechanism has just two internal states $A$ and $B$.
When the ratchet is not interacting with bits, the two states can have
different energies. We take $E_A=0$ and $E_B=-\alpha \kB T$, without loss of
generality. Since the bits interact with the ratchet one at a time, we only
need to specify the interaction energy of the ratchet and an individual bit.
The interaction energy is zero if the bit is in the state $Y = 0$, regardless
of the ratchet state, and it is $- \beta \kB T$ (or $+\beta \kB T$) if the bit
is in state $Y = 1$ and the ratchet is in state $A$ (or $B$). See Fig.
\ref{fig:Energies} for a graphical depiction of the energy scheme under
``Ratchet $\otimes$ Bit''.

The scheme is further modified by the interaction of the weight with the
ratchet and bit string.  We attach the weight to the ratchet-bit system such
that when the latter transitions from the $B \otimes 0$ state to the $A \otimes
1$ state it lifts the weight, doing a constant amount $w \kB T$ of work. As a
result, the energy of the composite system---Demon, interacting bit, and
weight---increases by $w \kB T$ whenever the transition $B \otimes 0
\rightarrow A \otimes 1$ takes place, the required energy being extracted from
the heat reservoir $Z_N$. The rightmost part of Fig.~\ref{fig:Energies}
indicates this by raising the energy level of $A \otimes 1$ by $w \kB T$
compared to its previous value. Since the transitions between $A \otimes 1$ and
$B \otimes 1$ do not involve the weight, their relative energy difference
remains unaffected. An increase in the energy of $A \otimes 1$ by $w \kB T$
therefore implies the same increase in the energy of $B \otimes 1$. Again, see
Fig. \ref{fig:Energies} for the energy scheme under ``Ratchet $\otimes$ Bit
$\otimes$ Weight''.

The time evolution over the joint state space of the ratchet, last bit of the
input string, and weight is governed by a Markov dynamic, specified by
state-transition matrix $M$. If, at the beginning of the $N$th interaction
interval at time $t=\tau (N-1)+0^+$, the ratchet is in state $X_N = x_N$ and the
input bit is in state $Y_N = y_N$, then let $M_{x_N \otimes y_N \rightarrow x_{N+1} \otimes
y'_{N} }$ be the probability $\Pr(x_{N+1},y'_{N}|x_N,y_N)$ that the ratchet is in state $X_N = x_{N+1}$
and the bit is in state $Y_N = y'_{N}$ at the end of the interaction interval
$t=\tau(N-1)+\tau^-$.  $X_N$ and $Y_N$ at the end of the $N$th interaction
interval become $X_{N+1}$ and $Y'_N$ respectively at the beginning of the
$N+1$th interaction interval.  Since we assume the system is thermalized with a
bath at temperature $T$, the ratchet dynamics obey detailed balance. And so,
transition rates are governed by the energy differences between joint states:
\begin{align}
\frac{M_{x_N \otimes y_N \rightarrow x_{N+1} \otimes y'_N }}{M_{x_{N+1} \otimes y'_N \rightarrow x_N \otimes y_N }}
  = e^{(E_{x_{N+1} \otimes y'_N}-E_{x_N\otimes y_N})/ \kB T}
  ~.
\end{align}

\begin{figure}[!ht]
\centering
\includegraphics[width=.5\columnwidth]{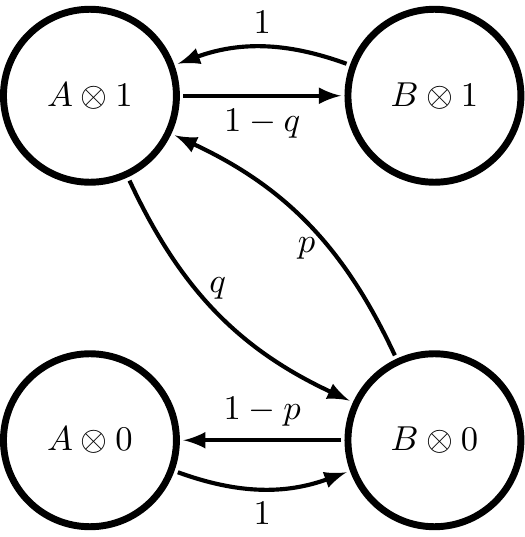}
\caption{The Markovian, detailed-balance dynamic over the joint states of the
  ratchet and interacting bit.
  }
\label{fig:MarkovModel}
\end{figure}

There is substantial flexibility in constructing a detailed-balanced Markov
dynamic for the ratchet, interaction bit, and weight. Consistent with our theme
of simplicity, we choose one that has only six allowed transitions: $A \otimes
0 \leftrightarrow B \otimes 0$, $A \otimes 1 \leftrightarrow B \otimes 1$, and
$A \otimes 1 \leftrightarrow B \otimes 0$. Such a model is convenient to
consider, since it can be described by just two transition probabilities $0\leq
p\leq 1$ and $0\leq q \leq 1$, as shown in Fig. \ref{fig:MarkovModel}.

\newcommand{\StateDist} { \mathbf{p} }

The Markov transition matrix for this system is given by:
\begin{align}
M = \left[ \begin{array}{cccc}
				0 & 1- p & 0 & 0 \\
				1 & 0 & q & 0 \\
				0 & p & 0 & 1 \\
				0 & 0 & 1-q & 0
			\end{array}\right].
\end{align}
This allows allows us to calculate the state distribution
$\StateDist((N-1)\tau+\tau^-)$ at the end of the $N$th interaction interval
from the state distribution $\StateDist((N-1)\tau+0^+)$ at the interval's
beginning via:
\begin{align}
\StateDist((N-1)\tau+ \tau^-) = M  \StateDist ((N-1)\tau+0^+)
  ~,
\end{align}
where the probability vector is indexed $\StateDist = (\Pr(A \otimes 0), \Pr(B \otimes 0),\Pr(A\otimes 1), \Pr(B \otimes 1))^\top$. To satisfy detailed
balance, we find that $\alpha$, $\beta$, and $w$ should be:
\begin{align}
\alpha  & = - \ln(1-p) \label{eq:alpha} ~,\\
  \beta & = -\frac{1}{2} \ln \left[(1-p)(1-q)\right]
     \label{eq:beta} ~,~\text{and}\\ 
      w & = \ln \left( \frac{q\sqrt{1-p}}{p \sqrt{1-q}}\right)
      \label{eq:w}
  ~.
\end{align}
(Appendix \ref{app:EnergyDesign} details the relationships between the transitions probabilities and energy levels.)

This simple model is particularly useful since, as we show shortly, it captures
the full range of thermodynamic functionality familiar from previous models
and, more importantly, it makes it possible to exactly calculate informational
properties of the output string analytically.

Now that we know how the ratchet interacts with the bit string and weight, we
need to characterize the input string to predict the energy flow through the
ratchet. As in the ratchet models of Refs. \cite{Mand012a,Lu14a}, we consider
an input generated by a biased coin---$\Pr(Y_N = 0) = b$ at each $N$---which
has no correlations between successive bits. For this input, the steady state
distributions at the beginning and end of the interaction interval $\tau$ are:
\begin{align}
\mathbf{p}^\text{s}(0^+) & = \frac{1}{2} \left[\begin{array}{c}
					b \\ 
					b \\
					1-b \\ 
					1-b
				\end{array}\right] ~\text{and} \nonumber \\
\mathbf{p}^\text{s}(\tau^-) & = \frac{1}{2} \left[ \begin{array}{c}
					b(1-p) \\
					b + q - bq \\
					bp + 1 - b \\
					(1-b)(1-q)
					\end{array} \right]
  	~.
\label{eq:ProbS}
\end{align} 
These distributions are needed to calculate the work done by the ratchet.

To calculate net extracted work by the ratchet we need to consider three
work-exchange steps for each interaction interval: (1) when the ratchet gets
attached to a new bit, to account for their interaction energy; (2) when the
joint transitions $B \otimes 0 \leftrightarrow A \otimes 1$ take place, to
account for the raising or lowering of the weight; and (3) when the ratchet
detaches itself from the old bit, again, to account for their nonzero
interaction energy. We refer to these incremental works as $W_1$, $ W_2$, and $
W_3$, respectively.

Consider the work $W_1$. If the new bit is in state $0$, from
Fig.~\ref{fig:Energies} we see that there is no change in the energy of the
joint system of the ratchet and the bit. However, if the new bit is $1$ and the
initial state of the ratchet is $A$, energy of the ratchet-bit joint system
decreases from 0 to $- \beta$. The corresponding energy is gained as work by
the mechanism that makes the ratchet move past the tape of bits. Similarly, if
the new bit is $1$ and the initial state of the ratchet is $B$, there is an
increase in the joint state energy by $\beta$; this amount of energy is now
taken away from the driving mechanism of the ratchet. In the steady state, the
average work gain $\langle W_1 \rangle$ is then obtained from the average
decrease in energy of the joint (ratchet-bit) system:
\begin{align}
\langle  W_1 \rangle
  & = - \sum_{ \substack{x \in \{A,B\} \\ y \in \{0,1\} } }
  p_{x \otimes y}^\text{s}(0^+)  \left( E_{x\otimes y} -E_x - E_y
  \right) \nonumber \\
  & = 0
  ~,
\end{align}
where we used the probabilities in Eq.~(\ref{eq:ProbS}) and Fig.~\ref{fig:Energies}'s energies.

By a similar argument, the average work $\langle W_3 \rangle$ is equal to the
average decrease in the energy of the joint system on the departure of the
ratchet, given by:
\begin{align}
\langle W_3 \rangle = - \frac{\kB T}{2} \beta [q + b (p-q)]
  ~.
\end{align}
Note that the cost of moving the Demon on the bit string (or moving the string
past a stationary Demon) is accounted for in works $W_1$ and $W_3$.

Work $W_2$ is associated with raising and lowering of the weight depicted in
Fig.~\ref{fig:FullRatchet}. Since transitions $B \otimes 0 \rightarrow A
\otimes 1$ raise the weight to give work $\kB T w$ and reverse transitions $B
\otimes 0 \leftarrow A \otimes 1$ lower the weight consuming equal amount of
work, the average work gain $\langle  W_2 \rangle$ must be $\kB T w$ times the
net probability transition along the former direction, which is $[ T_{B \otimes 0 \to A \otimes 1} p^\text{s}_{B \otimes 0}(0^+) 
  - T{_{A \otimes 1 \to A \otimes 1}} p^\text{s}_{A \otimes 1}(0^+)] $. This leads to the following expression: 
\begin{align}
\langle W_2 \rangle =  \frac{\kB T w}{2} [- q + b (p + q)]
  ~,
\end{align}
where we used the probabilities in Eq.~(\ref{eq:ProbS}). 

The total work supplied by the ratchet and a bit is their sum:
\begin{align}
\langle  W \rangle
  & = \langle W_1 \rangle + \langle W_2 \rangle
    + \langle W_3 \rangle
    \label{eq:WorkNet}
	\\
  & = \frac{\kB T}{2} [ (pb-q+qb) \ln{\left(\frac{q}{p}\right)} \nonumber \\
  & \quad\quad + (1-b)q \ln (1-q)+pb \ln (1-p) ]
  \nonumber
  ~.  
\end{align}
Note that we considered the total amount amount of work that can be gained by
the system, not just that obtained by raising the weight. Why? As we shall see
in Sec.~\ref{sec:TF}, the former is the thermodynamically more relevant
quantity. A similar energetic scheme that incorporates the effects of
interaction has also been discussed in Ref.~\cite{Um2015}.

In this way, we exactly calculated the work term in Eq.
(\ref{eq:NewSecondLaw}). We still need to calculate the entropy rate of the
output and input strings to validate the proposed Second Law. For this, we
introduce an information-theoretic formalism to monitor processing of the bit
strings by the ratchet.

\section{Information}

To analytically calculate the input and output entropy rates, we consider how
the strings are generated. A natural way to incorporate temporal correlations
in the input string is to model its generator by a finite-state hidden Markov
model (HMM), since HMMs are strictly more powerful than Markov chains in the
sense that finite-state HMMs can generate all processes produced by Markov
chains, but the reverse is not true. For example, there are processes generated
by finite HMMs that cannot be by any finite-state Markov chain. In short, HMMs
give a compact representations for a wider range of memoryful processes.

Consider possible input strings to the ratchet. With or without correlations
between bits, they can be described by an HMM generator with a finite set of,
say, $K$ states and a set of two symbol-labeled transition matrices $T^{(0)}$
and $T^{(1)}$, where:
\begin{align}
T^{(y_N)}_{s_N \to s_{N+1}} = \Pr(Y_N = y_N,S_{N+1}=s_{N+1}|S_N=s_N)
\end{align}
is the probability of outputting $y_N$ for the $N$th bit of the input string and
transitioning to internal state $s_{N+1}$ given that the HMM was in state $s_N$.

When it comes to the output string, in contrast, we have no choice. We are
forced to use HMMs. Since the current input bit state $Y_N$ and ratchet state
$X_N$ are not explicitly captured in the current output bit state $Y'_N$, $Y_N$
and $X_N$ are hidden variables. As we noted before, calculating HMM entropy
rates is a known challenging problem \cite{Blac57b,Marc11a}. Much of the
difficulty stems from the fact that in HMM-generated processes the effects of
internal states are only indirectly observed and, even then, appear only over
long output sequences.  

We can circumvent this difficulty by using \emph{unifilar} HMMs, in which the
current state and generated symbol uniquely determine the next state. This is
a key technical contribution here since for unifilar HMMs the entropy rate is
exactly calculable, as we now explain. Unifilar HMMs internal states are a
causal partitioning of the past, meaning that every past $w$ maps to a
particular state through some function $f$ and so:
\begin{align}
\Pr(Y_N = y_N|Y_{0:N}=w) = \Pr(Y_N = y_N|S_N=f(w))
  ~.
\end{align}
As a consequence, the entropy rate $\hmu$ in its block-entropy form (Eq.
(\ref{eq:EntropyRateViaBlocks})) can be re-expressed in terms of the transition
matrices. First, recall the alternative, equivalent form for entropy rate:
$\hmu = \lim_{N \rightarrow \infty} \H[Y_N|Y_{0:N}]$. Second, since $S_N$
captures all the dependence of $Y_N$ on the past, $\hmu = \lim_{N \rightarrow
\infty} \H[Y_N|S_N]$. This finally leads to a closed-form for the entropy rate
\cite{Crut01a}:
\begin{align}
\hmu & = \lim_{N \rightarrow \infty} \H[Y_N|S_N] \nonumber \\
  & = -\sum_{y_N,s_N,s_{N+1}} \pi_{s_N}
  T^{(y_N)}_{s_N\rightarrow s_{N+1}} \log_2 T^{(y_N)}_{s_N \rightarrow s_{N+1}}
  ~,
\label{eq:MachineEntropyRate}
\end{align}
where $\pi$ is the stationary distribution over the unifilar HMM's states.

\begin{figure}[!ht]
\centering
\includegraphics[width=.6\columnwidth]{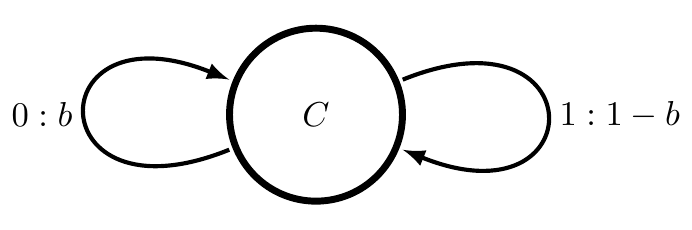}
\caption{Biased coin input string as a unifilar hidden Markov model with
  bias $\Pr(Y = 0) = b$.
  }
\label{fig:DrivingMachine}
\end{figure} 

Let's now put these observations to work. Here, we assume the ratchet's input string was generated by a memoryless biased coin. Figure \ref{fig:DrivingMachine} shows its (minimal-size) unifilar HMM.
The single internal state $C$ implies that the process is memoryless and the
bits are uncorrelated. The HMM's symbol-labeled ($1 \times 1$) transition
matrices are $T^{(0)}=\left[b\right]$ and $T^{(1)}=\left[1-b\right]$.
The transition from state $C$ to itself labeled $0:b$ means that if the system
is in state $C$, then it transitions to state $C$ and outputs $Y = 0$ with
probability $b$. Since this model is unifilar, we can calculate the input-string
entropy rate from Eq. (\ref{eq:MachineEntropyRate}) and see that it is the single-symbol entropy of bias $b$:
\begin{align}
\hmu & = \H(b) \nonumber \\
  & \equiv - b \log_2 b - (1-b) \log_2 (1-b)
  ~,
\end{align}
where $\H(b)$ is the (base $2$) binary entropy function \cite{Cove06a}.

The more challenging part of our overall analysis is to determine the entropy
rate of the output string. Even if the input is uncorrelated, it's possible
that the ratchet creates temporal correlations in the output string. (Indeed,
these correlations reflect the ratchet's operation and so its thermodynamic
behavior, as we shall see below.) To calculate the effect of these
correlations, we need a generating unifilar HMM for the output process---a
process produced by the ratchet being driven by the input.

When discussing the ratchet energetics, there was a Markov dynamic $M$ over the
ratchet-bit joint state space. Here, it is now controlled by bits from the
input string and writes the result of the thermal interaction with the ratchet
to the output string. In this way, $M$ becomes an input-output machine or
\emph{transducer} \cite{Barn13a}. In fact, this transducer is a communication
channel in the sense of Shannon \cite{Shan48a} that communicates the input bit
sequence to the output bit sequence. However, it is a channel with memory.  Its
internal states correspond to the ratchet's states. To work with $M$, we
rewrite it componentwise as:
\begin{align}
M^{(y'_N|y_N)}_{x_N \rightarrow x_{N+1}} = M_{x_N \otimes y_N \rightarrow x_{N+1} \otimes y'_N}
\end{align}
to evoke its re-tooled operation. The probability of generating bit $y'_N$ and
transitioning to ratchet state $x_{N+1}$, given that the input bit is $y_N$ and
the ratchet is in state $x_N$, is:
\begin{align}
& M^{(y'_N|y_N)}_{x_N \rightarrow x_{N+1}} = \\
   & \quad\quad \Pr(Y'_N=y'_N,X_{N+1}=x_{N+1}|Y_N=y_N,X_N=x_N)
  \nonumber ~.
\end{align}
This allows us to exactly calculate the symbol-labeled transition matrices,
$T'^{(0)}$ and $T'^{(1)}$, of the HMM that generates the output string:
\begin{align}
T'^{(y'_N)}_{s_N \otimes x_N \rightarrow s_{N+1} \otimes x_{N+1} }
  = \sum_{y_N}M^{(y'_N|y_N)}_{x_N \rightarrow x_{N+1}}T^{(y_N)}_{s_N \rightarrow s_{N+1}}
  ~.
\end{align}
The joint states of the ratchet and the internal states of the input process
are the internal states of the output HMM, with $x_N,x_{N+1}  \in \{A, B\}$ and $s_N, s_{N+1}
\in \{C\}$ in the present case. This approach is a powerful tool for directly
analyzing informational properties of the output process.

By adopting the transducer perspective, it is possible to find HMMs for the
output processes of previous ratchet models, such as in Refs.
\cite{Mand012a,Lu14a}. However, their generating HMMs are highly nonunifilar,
meaning that knowing the current internal state and output allows for many
alternative internal-state paths. And, this precludes writing down closed-form
expressions for informational quantities, as we do here. Said simply, the
essential problem is that those models build in too many transitions.
Ameliorating this constraint led to the Markov dynamic shown in Fig.
\ref{fig:MarkovModel} with two ratchet states and sparse transitions. Although
this ratchet's behavior cannot be produced by a rate equation, due to the
limited transitions, it respects detailed balance.

\begin{figure}[!ht]
\centering
\includegraphics[width=.5\columnwidth]{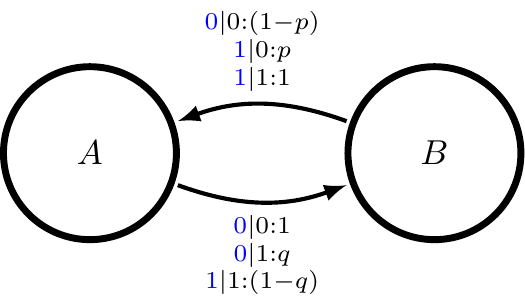}
\caption{The Maxwellian ratchet's transducer.}
\label{fig:Transducer}
\end{figure}

Figure \ref{fig:Transducer} shows our two-state ratchet's transducer. As noted
above, it's internal states are the ratchet states. Each transition is labeled
$\textcolor{blue}{y'}|y:p$, where $\textcolor{blue}{y'}$ is the output,
conditioned on an input $y$, with probability $p$.

\begin{figure}[!ht]
\centering
\includegraphics[width=.5\columnwidth]{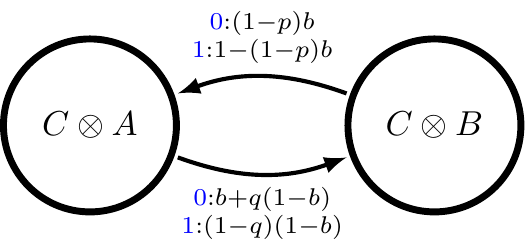}
\caption{Unifilar HMM for the output string generated by the ratchet driven
  by a coin with bias $b$.
  }
\label{fig:Outputs}
\end{figure}

We can drive this ratchet (transducer) with any input, but for comparison with
previous work, we drive it with the memoryless biased coin process just
introduced and shown in Fig. \ref{fig:DrivingMachine}. The resulting unifilar
HMM for the output string is shown in Fig. \ref{fig:Outputs}. The corresponding
symbol-labeled transition matrices are:
\begin{align}
T'^{(0)} & = \left[ \begin{array}{cccc}
				0 & (1-p)b   \\
				b+q(1-b) & 0
			\end{array}\right]
      ~,~ \text{and} \\
T'^{(1)} & = \left[ \begin{array}{cccc}
				0 & 1-(1-p)b  \\
				(1-q)(1-b)  & 0
			\end{array}\right]
  	~.
\label{eq:OutputProcess}
\end{align}

Using these we can complete our validation of the proposed Second Law, by
exactly calculating the entropy rate of the output string. We find:
\begin{align}
\hmu' & = \lim_{N\rightarrow \infty} \H[Y'_N|Y'_{0:N}] \nonumber \\
   & = \lim_{N\rightarrow \infty} \H[Y'_N|S_N] \nonumber \\
   & = \frac{\H(b(1-p))}{2}+\frac{\H((1-b)(1-q))}{2}
   ~.
\end{align}
We note that this is less than or equal to the
(unconditioned) single-symbol
entropy for the output process:
\begin{align}
\hmu' & \leq \H[Y'_0] \nonumber \\
  & = \H \left( (b(1-p) +(1-b)(1-q)) / 2 \right)
  ~.
\end{align}
Any difference between $\hmu'$ and single-symbol entropy $\H[Y_{0}]$
indicates correlations that the ratchet created in the output from the
uncorrelated input string. In short, the entropy rate gives a more accurate
picture of how information is flowing between bit strings and the heat bath.
And, as we now demonstrate, the entropy rate leads to correctly identifying
important classes of ratchet thermodynamic functioning---functionality the single-symbol entropy misses.

\section{Thermodynamic Functionality}
\label{sec:TF}

Let's step back to review and set context for exploring the ratchet's
thermodynamic functionality as we vary its parameters. Our main results
are analytical, provided in closed-form. First, we derived a modified version
of the Second Law of Thermodynamics for information ratchets in terms of the
difference between the Kolmogorov-Sinai entropy of the input and output
strings:
\begin{align}
\langle  W \rangle \leq \kB T  \ln2 \, \Delta \hmu
  ~,
\end{align}
where $\Delta \hmu = \hmu' - \hmu$. The improvement here takes into account
correlations within the input string and those in the output string actively
generated by the ratchet during its operation. From basic information-theoretic
identities we know this bound is stricter for memoryless inputs than previous
relations \cite{Jarz97a} that ignored correlations. However, by how much? And,
this brings us to our second main result. We gave analytic expressions for both
the input and output entropy rates and the work done by the Demon. Now, we are
ready to test that the bound is satisfied and to see how much stricter it is
than earlier approximations.

\begin{figure}[!ht]
\centering
\includegraphics[width=\columnwidth]{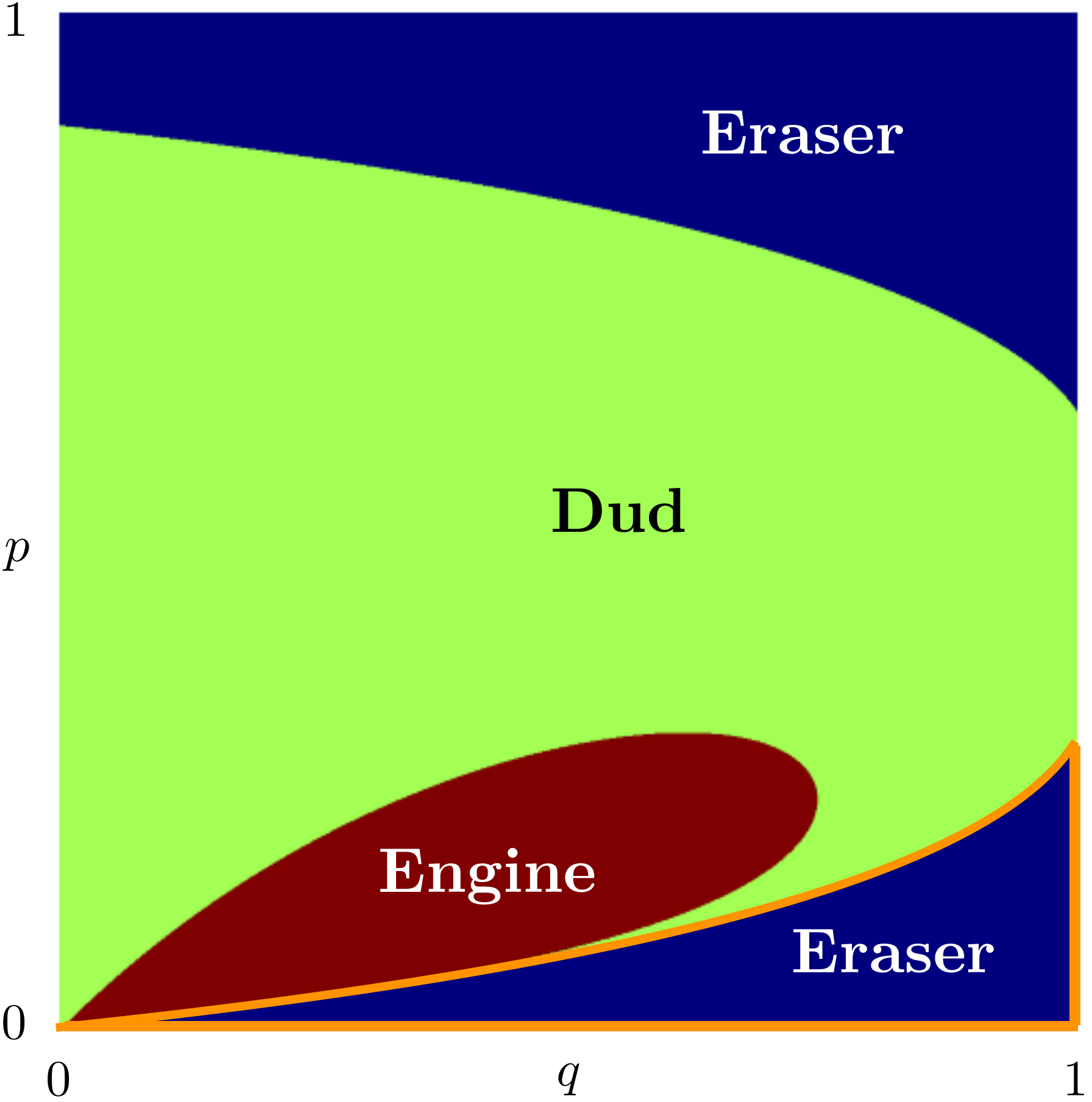}
\caption{Information ratchet thermodynamic functionality at input bias $b=0.9$:
  Engine: $(p,q)$ such that
  $0 < \langle W \rangle \leq \kB T \ln 2 \, \Delta h_\mu$. 
  Eraser: $(p,q)$ such that
  $\langle W \rangle \leq \kB T \ln 2 \, \Delta h_\mu < 0$.
  Dud: $(p,q)$ such that
  $\langle W \rangle \leq 0 \leq \kB T \ln 2 \, \Delta h_\mu$. 
  }
\label{fig:RatchetBehavior}
\end{figure}

We find diverse thermodynamic behaviors as shown in Figure \ref{fig:RatchetBehavior}, which describes ratchet thermodynamic function at input bias $b = 0.9$. We note that there are analogous behaviors for all values of input bias.
We identified three possible behaviors for the ratchet: \emph{Engine},
\emph{Dud}, and \emph{Eraser}. Nowhere does the ratchet violate the rule
$\langle W \rangle \leq \kB T \ln 2 \, \Delta h_\mu$. The engine
regime is defined by $(p,q)$ for which $\kB T \ln 2 \, \Delta h_\mu \geq
\langle  W \rangle >0$ since work is positive.  This is the only
condition for which the ratchet extracts work. The eraser regime is defined by
$0> \kB T \ln 2 \, \Delta h_\mu \geq \langle  W \rangle$, meaning that
work is extracted from the work reservoir while the uncertainty in the bit
string decreases. In the dud regime, those $(p,q)$ for which $\kB T \ln 2 \,
\Delta h_\mu \geq 0 \geq \langle W \rangle $, the ratchet is neither
able to erase information nor is it able to do useful work.

\begin{figure}[!ht]
\centering
\includegraphics[width=\columnwidth]{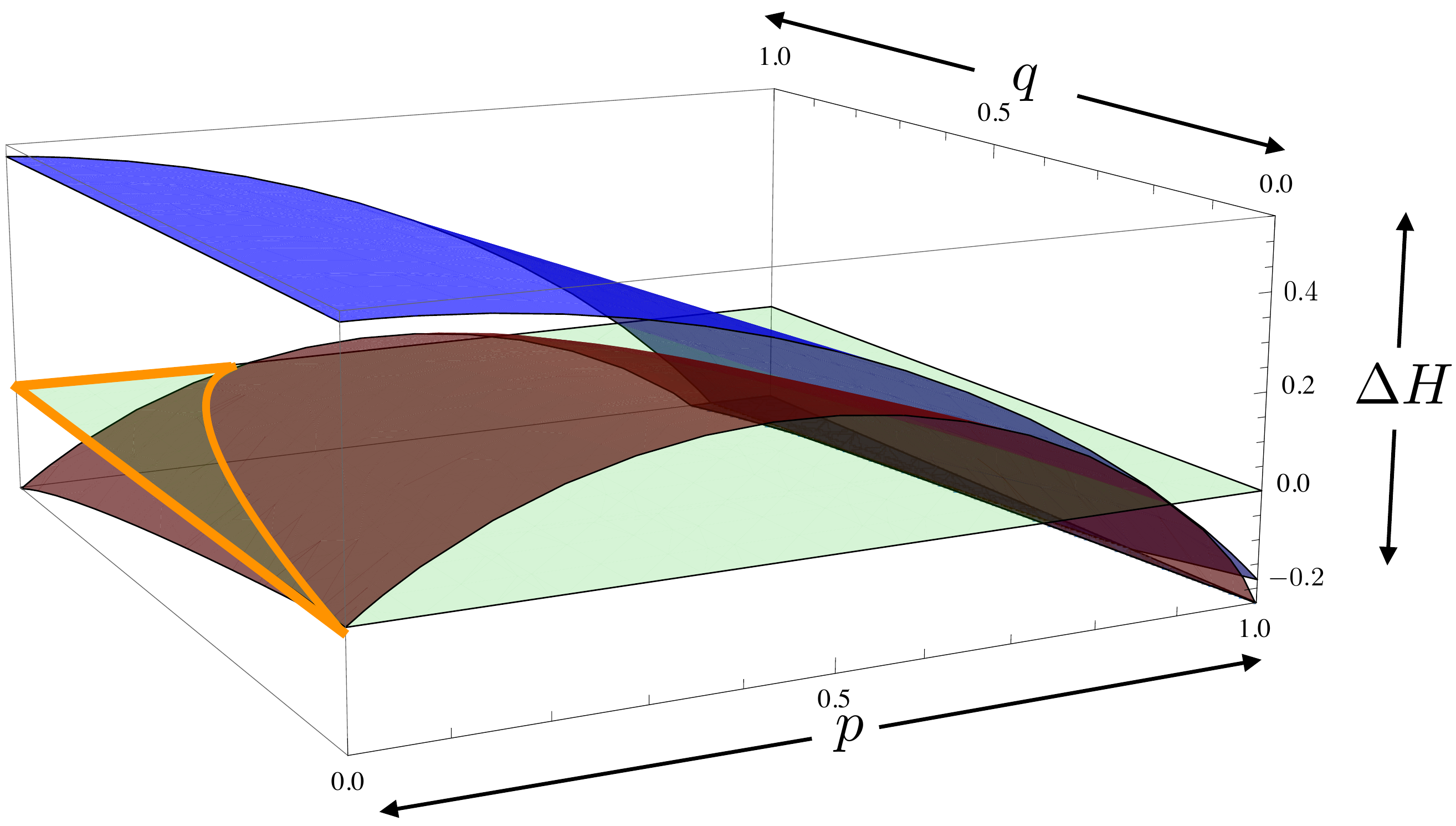}
\caption{Exact entropy rate difference $\Delta \hmu$ (red) is a much stricter
  bound on work than the difference in single-symbol entropy $\Delta \H[Y_{0}]$
  (blue). The zero surface (light green) highlights where both entropies are
  greater than zero and so is an aid to identifying functionalities.
  }
\label{fig:EntropyRates}
\end{figure}

At first blush, these are the same behavior types reported by Ref. \cite{Mand012a}, except that
we have stronger bounds on the work now with $\kB T \ln 2 \, \Delta \hmu$,
compared to the single-symbol entropy approximation. The stricter bound gives
deeper insight into ratchet functionality. To give a concrete comparison,
Fig. \ref{fig:EntropyRates} plots the single-symbol entropy difference $\Delta
\H[Y_{0}]$ and the entropy rate difference $\Delta \hmu$, with a flat surface
identifying zero entropy change, for all $p$ and $q$ and at $b = 0.9$.

In the present setting where input symbols are uncorrelated, the blue $\Delta
\!\H[Y_{0}]$ surface lies above the red $\Delta \hmu$ surface for all parameters,
confirming that the single-symbol entropy difference is always greater than the
entropy rate difference. It should also be noted for this choice of input bias
$b$ and for larger $p$, $\Delta\!\H[Y_{0}]$ and $\Delta \hmu$ are close, but they
diverge for smaller $p$. They diverge so much, however, that looking only at
single-symbol entropy approximation misses an entire low-$p$ region,
highlighted in orange in Fig. \ref{fig:EntropyRates} and
\ref{fig:RatchetBehavior}, where $\Delta h_\mu$ dips below zero and the ratchet
functions as eraser.

The orange-outlined low-$p$ erasure region is particularly interesting, as it
hosts a new functionality not previously identified: The ratchet removes
multiple-bit uncertainty, effectively erasing incoming bits by adding temporal
order, all the while increasing the uncertainty in individual incoming bits.
The existence of this mode of erasure is highly counterintuitive in light of
the fact the Demon interacts with only one bit at a time. In contrast,
operation in the erasure region at high $p$, like that in previous Demons,
simply reduces single-bit uncertainty. Moreover, the low-$p$ erasure region
lies very close to the region where ratchet functions as an engine, as shown in
Fig. \ref{fig:RatchetBehavior}. As one approaches $(p,q) = (0,0)$ the eraser
and engine regions become arbitrarily close in parameter space. This is a
functionally meaningful region, since the device can be easily and efficiently
switched between distinct modalities---an eraser or an engine.

In contrast, without knowing the exact entropy rate, it appears that the
engine region of the ratchet's parameter space is isolated from the eraser
region by a large dud region and that the ratchet is not tunable. Thus, knowing
the correlations between bits in the output string allows one to predict
additional functionality that otherwise is obscured when one only considers the
single-symbol entropy of the output string.

As alluded to above, we can also consider structured input strings generated by
memoryful processes, unlike the memoryless biased coin. While correlations in
the output string are relevant to the energetic behavior of this ratchet, it
turns out that input string correlations are not. The work done by the ratchet
depends only on the input's single-symbol bias $b$.  That said, elsewhere we
will explore more intelligent ratchets that take advantage of input string
correlations to do additional work.

\section*{Conclusion}

Thermodynamic systems that include information reservoirs as well as thermal
and work reservoirs are an area of growing interest, driven in many cases by
biomolecular chemistry or nanoscale physics and engineering. With the ability
to manipulate thermal systems on energy scales closer and closer to the level
of thermal fluctuations $\kB T$, information becomes critical to the flow of
energy. Our model of a ratchet and a bit string as the information reservoir is
very flexible and our methods showed how to analyze a broad class of such
controlled thermodynamic systems. Central to identifying thermodynamic
functionality was our deriving Eq. (\ref{eq:NewSecondLaw}), based on the control
system's Kolmogorov-Sinai entropy, that holds in all situations of memoryful
or memoryless ratchets and correlated or uncorrelated input processes and
that typically provides the tightest quantitative bound on work. This
improvement comes directly from tracking Demon information production over
system trajectories, not from time-local, configurational entropies.

Though its perspective and methods were not explicitly highlighted,
\emph{computational mechanics} \cite{Crut12a} played a critical role in the
foregoing analyses, from its focus on structure and calculating all system
component correlations to the technical emphasis on unifilarity in Demon
models.  Its full impact was not fully explicated here and is left to sequels
and sister works. Two complementary computational mechanics analyses of
information engines come to mind, in this light.  The first is Ref.
\cite{Boyd14b}'s demonstration that the chaotic instability in Szilard's
Engine, reconceived as a deterministic dynamical system, is key to its ability
to extract heat from a reservoir. This, too, highlights the role of
Kolmogorov-Sinai dynamical entropy. Another is the thorough-going extension of
fluctuation relations to show how intelligent agents can harvest energy when
synchronizing to the fluctuations from a structured environment \cite{Riec15a}.

This is to say, in effect, the foregoing showed that computational mechanics is
a natural framework for analyzing a ratchet interacting with an information
reservoir to extract work from a thermal bath. The input and output strings
that compose the information reservoir are best described by unifilar HMM
generators, since they allow for exact calculation of any informational
property of the strings, most importantly the entropy rate. In fact, the
control system components are the \eMs\ and \eTs\ of computational mechanics
\cite{Crut12a,Barn13a}.

By allowing one to exactly calculate the asymptotic entropy rate, we identified
more functionality in the effective thermodynamic \eTs\ than previous methods
can reveal. Two immediate consequences were that we identified a new kind of
thermodynamic eraser and found that our ratchet is easily tunable between an
eraser and an engine---functionalities suggesting that real-world ratchets
exhibit memory to take advantage of correlated environmental fluctuations, as
well as hinting at useful future engineering applications.


\section*{Acknowledgments}

We thank M. DeWeese and S. Marzen for useful conversations. As an External Faculty member, JPC thanks the Santa Fe Institute for its hospitality during visits. This work was supported in part by the U. S. Army Research Laboratory and the U. S. Army Research Office under contracts W911NF-13-1-0390 and W911NF-12-1-0234.

\appendix

\section{Derivation of Eq. (\ref{eq:NewSecondLaw})}
\label{app:NewSecondLaw}

Here, we reframe the Second Law of Thermodynamics, deriving an expression of it
that makes only one assumption about the information ratchet operating along
the bit string: the ratchet accesses only a finite number of internal states.
This constraint is rather mild and, thus, the bounds on thermodynamic
functioning derived from the new Second Law apply quite broadly.

The original Second Law of Thermodynamics states that the total change in
entropy of an isolated system must be nonnegative over any
time interval. By considering a system composed of a thermal reservoir,
information reservoir, and ratchet, in the following we derive an analog in
terms of rates, rather than total configurational entropy changes.

Due to the Second Law, we insist that the change in thermodynamic entropy of
the closed system is positive for any number $N$ of time steps. If $X$ denotes
the ratchet, $Y$ the bit string, and $Z$ the heat bath, this assumption
translates to:
\begin{align}
\triangle S[X,Y,Z] \geq 0
  ~.
\end{align}
Note that we do not include a term for the weight (a mechanical energy reservoir), since it does not contribute
to the thermodynamic entropy. Expressing the thermodynamic entropy $S$ in terms
the Shannon entropy of the random variables $S[X,Y,Z] = \kB \ln 2 \H[X,Y,Z]$,
we have the condition:
\begin{align}
\triangle \!\H[X,Y,Z]\geq 0
  ~.
\end{align}

To be more precise, this is true over any number of time steps $N$. If we have
our system $X$, we denote the random variable for its state at time step $N$
by $X_N$. The information reservoir $Y$ is a semi-infinite string. At time zero, the
string is composed entirely of the bits of the input process, for which the
random variable is denoted $Y_{0:\infty}$. The ratchet transduces these inputs,
starting with $Y_0$ and generating the output bit string, the entirety of which
is expressed by the random variable $Y'_{0: \infty}$. At the $N$th time step,
the first $N$ bits of the input $Y$ have been converted into the first $N$ bits of
the output $Y^\prime$, so the random variable for the input-output bit string is
$Y_{N:\infty} \otimes Y'_{0:N}$.  Thus, the change in entropy from the initial time to the $N$th time step is:
\begin{align}
\triangle \H_N[X,Y,Z] &= \H[X_N,Y_{N:\infty}, Y'_{0:N},Z_N] \nonumber \\
  & \quad\quad - \H[X_0,Y_{0:\infty},Z_0] \\
  &= \H[X_N,Y_{N:\infty}, Y'_{0:N}]+ \H[Z_N] \nonumber \\
  & \quad\quad - \I[X_N,Y_{N:\infty}, Y'_{0:N};Z_N] \nonumber \\
  & \quad\quad - \H[X_0,Y_{0:\infty}]- \H[Z_0] \nonumber \\
  & \quad\quad + \I[X_0,Y_{0:\infty};Z_0]
  ~.
\end{align}
Note that the internal states of an infinite heat bath do not correlate with
the environment, since they have no memory of the environment. This means the
mutual informations $\I[X_N,Y_{N:\infty}, Y'_{0:N};Z_N]$ and
$\I[X_0,Y_{0:\infty};Z_0]$ of the thermal reservoir $Z$ with the bit string $Y$
and ratchet $X$ vanish. Also, note that the change in thermal bath entropy can
be expressed in terms of the heat dissipated $Q_N$ over the $N$ time steps:
\begin{align}
\triangle \!\H[Z] & = \H[Z_N] - \H[Z_0] \nonumber \\
  & = Q_N/\kB T \ln 2
  ~.
\end{align}
Thus, the Second Law naturally separates into energetic terms describing the
change in the heat bath and information terms describing the ratchet and bit
strings:
\begin{align}
\triangle \H_N[X,Y,Z] & = \frac{Q_N}{k_BT \, \ln2} \\
  & \quad\quad + \H[X_N,Y_{N:\infty},Y'_{0:N}] - \H[X_0,Y_{0:\infty}]
  \nonumber ~.
\end{align}
Since $\triangle \H \geq 0$, we can rewrite this as an entirely general lower
bound on the dissipated heat over a length $N \tau$ time interval, recalling
that $\tau$ is the ratchet-bit interaction time:
\begin{align}
Q_N \geq k_B T \, \ln 2
  \left( \H[X_0,Y_{0:\infty}] - \H[X_N,Y_{N:\infty},Y'_{0:N}] \right)
  ~.
\end{align}
This bound is superficially similar to Eq. (\ref{eq:NotFundamental}), but it's
true in all cases, as we have not yet made any assumptions about the ratchet.
However, its informational quantities are difficult to calculate for large
$N$ and, in their current form, do not give much insight. Thus, we look at the
infinite-time limit in order tease out hidden properties.

Over a time interval $N \tau$, the average heat dissipated per ratchet cycle is
$Q_N /N$. When we classify an engine's operation, we usually quantify
energy flows that neglect transient dynamics. These are just the heat
dissipated per cycle over infinite time $ \langle Q \rangle= \lim_{N\rightarrow
\infty} Q_N /N$, which has the lower bound:
\begin{align}
\langle Q \rangle \geq \lim_{N\rightarrow \infty} k_B T \, \ln 2 \frac{H[X_0,Y_{0:\infty}]-H[X_N,Y_{N:\infty},Y'_{0:N}]}{N}.
\end{align}

Assuming the ratchet has a finite number of internal states, each with finite
energy, then the bound can be simplified and written in terms of work. In this
case, the average work done is the opposite of the average dissipated heat:
$\langle W \rangle =-\langle Q \rangle$. And so, it has the upper bound:
\begin{align}
\langle W \rangle \leq  k_B T \, \ln 2 & \lim_{N\rightarrow \infty}
  \bigg(  \frac{H[Y_{N:\infty},Y'_{0:N}]-H[Y_{0:\infty}]}{N} \nonumber \\
  & +\frac{H[X_N]-H[X_0]}{N} \\
  & +\frac{I[X_0;Y_{0:\infty}]-I[X_N;Y_{N:\infty},Y'_{0:N}]}{N} \bigg)
  \nonumber ,
\end{align}
where the joint entropies are expanded in terms of their single-variable
entropies and mutual informations.

The entropies over the initial $X_0$ and final $X_N$ ratchet state
distributions monitor the change in ratchet memory---time-dependent versions of
its statistical complexity $\Cmu(N) = \H[X_N]$ \cite{Crut12a}. This time
dependence can be used to monitor how and when the ratchet synchronizes to the
incoming sequence, recognizing a sequence's temporal correlations. However,
since we assumed that the ratchet has finite states, the ratchet state-entropy
and also mutual information terms involving it are bounded above by the
logarithm of the number states. And so, they go to zero as $N \to \infty$, leaving the expression:
\begin{align}
\langle W \rangle &
  \leq  k_B T \, \ln 2 \lim_{N\rightarrow \infty}
  \bigg(  \frac{H[Y_{N:\infty},Y'_{0:N}]-H[Y_{0:\infty}]}{N}\bigg)
  ~.
\end{align}
With this, we have a very general upper bound for the work done by the ratchet in terms of just the input and output string variables.

Once again, we split the joint entropy term into it's components:
\begin{align}
\langle W \rangle \leq k_B T \, \ln 2
  \lim_{N\rightarrow \infty}
  & \bigg(  \frac{H[Y_{N:\infty}]-H[Y_{0:\infty}]}{N} \\
  & \quad + \frac{H[Y'_{0:N}]}{N}-\frac{I[Y_{N:\infty};Y'_{0:N}]}{N}\bigg)
  \nonumber ~.
\end{align}
In this we identify the output process's entropy rate $\hmu' = \lim_{N \rightarrow \infty} \H[Y'_{0:N}] / N$. While $\lim_{N \rightarrow \infty} \big(
\H[Y_{N:\infty}]- \H[Y_{0:\infty}] \big) / N$ looks unfamiliar, it is actually the
negative entropy rate $\hmu$ of the input process, so we find that:
\begin{align}
\langle W \rangle &  \leq  k_B T \, \ln 2  \bigg(  h'_\mu-h_\mu
 -\lim_{N\rightarrow \infty}\frac{I[Y_{N:\infty};Y'_{0:N}]}{N}\bigg).
\end{align}

\begin{figure}[tb]
\centering
\includegraphics[width=\columnwidth]{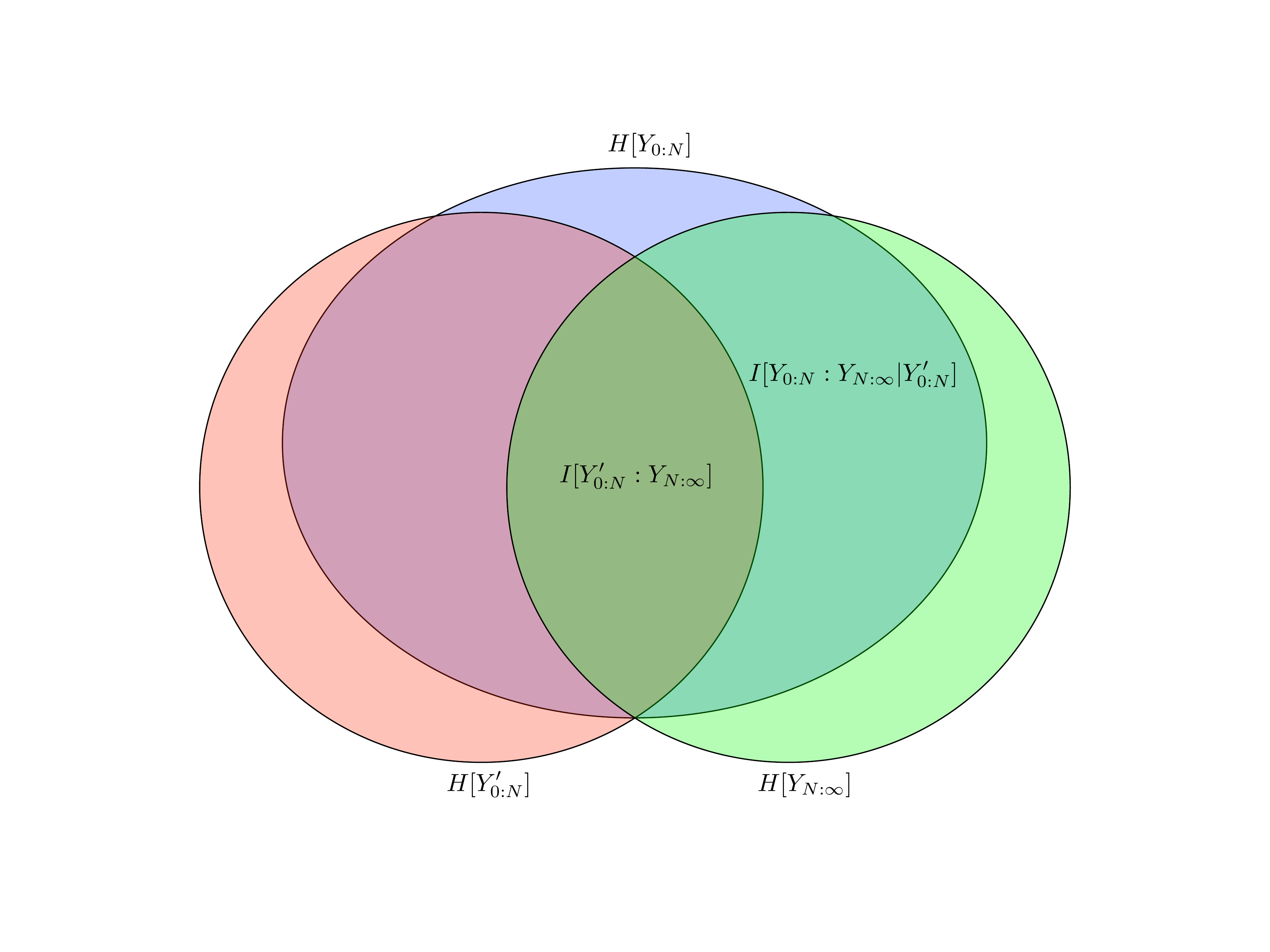}
\caption{The $N$ most recent variables of the input process shield the $N$
  variables of output from the rest of the input variables.
  }
\label{fig:Shielding}
\end{figure}

To understand the mutual information term, note that $Y'_{0:N}$ is generated from
$Y_{0:N}$, so it is independent of $Y_{N:\infty}$ conditioned on $Y_{0:N}$.
Essentially, $Y_{0:N}$ causally shields $Y'_{0:N}$ from $Y_{N:\infty}$, as
shown in information diagram \cite{Yeun08a} of Fig \ref{fig:Shielding}. This means:
\begin{align}
\I[Y_{N:\infty};Y'_{0:N}]
  = \I[Y_{N:\infty};Y_{0:N}]-\I[Y_{N:\infty};Y_{0:N}|Y'_{0:N}]
  ~.
\end{align}
This, in turn, gives: $\I[Y_{N:\infty};Y_{0:N}] \geq \I[Y_{N:\infty};Y'_{0:N}]
\geq 0$. Thus, we find the input process's excess entropy $\EE$ \cite{Crut01a}:
\begin{align}
\lim_{N \rightarrow \infty} \I[Y_{N:\infty};Y'_{0:N}]
  & \leq \lim_{N \rightarrow \infty} \I[Y_{N:\infty};Y_{0:N}] \nonumber \\
  & = \EE
  ~.
\end{align}
However, dividing by $N$ it's contribution vanishes:
\begin{align}
\lim_{N \rightarrow \infty} \frac{I[Y_{N: \infty};Y_{0:N}]}{N}
  & =  \lim_{N \rightarrow \infty}
  \left( \frac{H[Y_{0:N}]}{N}
  - \frac{H[Y_{0:N}|Y_{N: \infty}]}{N} \right) \nonumber \\
  & = \hmu- \hmu \nonumber \\
  & = 0
  ~.
\end{align}
Thus, we are left with the inequality of Eq. (\ref{eq:NewSecondLaw}):
\begin{align}
\langle W \rangle  \leq  k_B T \, \ln 2 \left(  \hmu' - \hmu \right)
  ~ ;
\label{eq:DerivedSecondLaw}
\end{align}
derived with minimal assumptions. Also, the appearance of
the statistical complexity and excess entropy, whose contributions this
particular derivation shows are asymptotically small, does indicate the
potential role of correlations in the input for finite time---times during
which the ratchet synchronizes to the incoming information \cite{Crut10a}.

One key difference between Eq. (\ref{eq:DerivedSecondLaw}) (equivalently, Eq.
(\ref{eq:NewSecondLaw})) and the more commonly used bound in Eq.
(\ref{eq:NotFundamental}), with the change in single-variable configurational
entropy $\H[Y'_{0}]-\H[Y_{0}]$, is that the former bound is true for all finite
ratchets and takes into account the production of information over time via the
Kolmogorov-Sinai entropies $\hmu$ and $\hmu'$. More generally, we do not look
at single-step changes in configurational
entropies---$\H[X_{N-1},Y_{N-1},Z_{N-1}] \to \H[X_N,Y_N,Z_N]$---but rather the
rate of production of information $\H[W_N | \ldots W_{N-2}, W_{N-1} ]$, where
$W_N = (X_N,Y_N,Z_N)$. This global dynamical entropy rate has contributions
from output rate $\hmu'$ and input rate $\hmu$.  This again indicates how Eq.
(\ref{eq:NotFundamental}) approximates Eq. (\ref{eq:DerivedSecondLaw}).

There are several special cases where the single-variable bound of Eq.
(\ref{eq:NotFundamental}) applies. In the case where the input is uncorrelated,
it holds, but it is a weaker bound than Eq. (\ref{eq:NewSecondLaw}) using
entropy rates. Also, in the case when the ratchet has no internal states and so
is memoryless, Eq. (\ref{eq:NotFundamental}) is satisfied. Interestingly, either
it or Eq. (\ref{eq:DerivedSecondLaw}) can be quantitatively stricter in this special case. However, in the most general case where the inputs are correlated
and the ratchet has memory, the bound using single-variable entropy is
incorrect, since there are cases where it is violated \cite{Mand15a}. Finally,
when the input-bit-ratchet interaction time $\tau$ grows the ratchet spends
much time thermalizing. The result is that the output string becomes
uncorrelated with the input and so the ratchet is effectively memoryless.
Whether by assumption or if it arises as the effective behavior, whenever the
ratchet is memoryless, it is ignorant of temporal correlations and so it and
the single-symbol entropy bounds are of limited physical import. These issues
will be discussed in detail in future works, but as a preview see Ref. \cite{Mand15a}.

\section{Designing Ratchet Energetics}
\label{app:EnergyDesign}

Figure~\ref{fig:MarkovModel} is one of the simplest information transducers for
which the outcomes are unifilar for uncorrelated inputs, resulting in the fact
that the correlations in the outgoing bits can be explicitly calculated. As
this calculation was a primary motivation in our work, we introduced the model
in Fig.~\ref{fig:MarkovModel} first and, only then, introduced the associated
energetic and thermodynamic quantities, as in Fig.~\ref{fig:Energies}. The
introduction of energetic and thermodynamic quantities for an abstract
transducer (as in Fig.~\ref{fig:MarkovModel}), however, is not trivial. Given a
transducer topology (such as the reverse ``Z" shape of the current model),
there are multiple possible energy schemes of which only a fraction are
consistent with all possible values of the associated transition probabilities.
However, more than one scheme is generally possible.

To show that only a fraction of all possible energetic schemes are consistent
with all possible parameter values, consider the case where the interaction
energy between the ratchet and a bit is zero, as in Ref.~\cite{Mand012a}. In
our model, this implies $\beta = 0$, or equivalently, $p = q = 0$ (from
Eq.~(\ref{eq:beta})). In other words, we cannot describe our model, valid for all values $0 < p, q < 1$, by the energy scheme in Fig.~\ref{fig:Energies} with $\beta = 0$. This is despite the fact that we have two other independent parameters $\alpha$ and $w$.


To show that, nonetheless, more than one scheme is possible, imagine the case
with $\alpha = \beta = 0$. Instead of just one mass, consider three masses such
that, whenever the transitions $A \otimes 0 \rightarrow B \otimes 0$, $B \otimes
0 \rightarrow A \otimes 1$, and $A \otimes 1 \rightarrow B \otimes 1$ take
place, we get works $\kB T \widetilde{W}_1$, $\kB T \widetilde{W}_2$, and $\kB T
\widetilde{W}_3$, respectively. We lose the corresponding amounts of work for the reverse transitions. This picture is consistent with the abstract model of Fig.~\ref{fig:MarkovModel} if the following requirements of detailed balance are satisfied:
\begin{align}
\label{eq:DBApp11}
\frac{1}{1 - p} & = \frac{M_{A \otimes 0 \rightarrow B \otimes 0}}{M_{B \otimes 0 \rightarrow A \otimes 0}} = e^{- \widetilde{W}_1} ~,\\
\label{eq:DBApp12}
\frac{p}{q} & = \frac{M_{B \otimes 0 \rightarrow A \otimes 1}}{M_{A \otimes 1
\rightarrow B \otimes 0}} = e^{- \widetilde{W}_2} ~,~\text{and} \\
\label{eq:DBApp13}
1-q & = \frac{M_{A \otimes 1 \rightarrow B \otimes 1}}{M_{B \otimes 1
\rightarrow A \otimes 1}} = e^{- \widetilde{W}_3}
  ~.
\end{align}
Existence of such an alternative scheme illustrates the fact
that given the abstract model of Fig.~\ref{fig:MarkovModel}, there is more than
one possible consistent energy scheme. We suggest that this will allow for
future engineering flexibility.

\bibliography{chaos}

\end{document}